\documentclass[%
 amsmath,amssymb,
 reprint,%
]{revtex4-1}

\usepackage{graphicx}
\usepackage{dcolumn}
\usepackage{bm}

\usepackage[utf8]{inputenc}
\usepackage[T1]{fontenc}
\usepackage{mathptmx}
\usepackage{etoolbox}
\DeclareUnicodeCharacter{0301}{\'{e}}

\makeatletter
\def\@email#1#2{%
 \endgroup
 \patchcmd{\titleblock@produce}
  {\frontmatter@RRAPformat}
  {\frontmatter@RRAPformat{\produce@RRAP{*#1\href{mailto:#2}{#2}}}\frontmatter@RRAPformat}
  {}{}
}%
\makeatother
\begin{document}


\title{Pair distribution function analysis for oxide defect identification through feature extraction and supervised learning}

\author{Shuyan Zhang$^1$}
\author{Jie Gong$^1$}%
\author{Daniel Xiao$^2$}
\author{B. Reeja Jayan$^1$}
\author{Alan J. H. McGaughey$^{1}$}\email{mcgaughey@cmu.edu.}
\affiliation{%
 $^1$Department of Mechanical Engineering, Carnegie Mellon University, \\Pittsburgh, Pennsylvania 15213, USA  
}%
\affiliation{%
 $^2$Department of Materials Science and Engineering, Carnegie Mellon University, \\Pittsburgh, Pennsylvania 15213, USA  
}%

\begin{abstract}
Feature extraction and a neural network model are applied to predict the defect types and concentrations in experimental TiO$_2$ samples. A dataset of TiO$_2$ structures with vacancies and interstitials of oxygen and titanium is built and the structures are relaxed using energy minimization. The features of the calculated pair distribution functions (PDFs) of these defected structures are extracted using linear methods (principal component analysis, non-negative matrix factorization) and non-linear methods (autoencoder, convolutional neural network). The extracted features are used as the inputs to a neural network that maps the feature weights to the concentration of each defect type. The performance of this machine learning pipeline is validated by predicting the defect concentrations based on experimentally-measured TiO$_2$ PDFs and comparing the results to brute-force predictions. A physics-based initialization of the autoencoder has the highest accuracy in predicting the defect concentrations. This model incorporates physical interpretability and predictability of material properties, enabling a more efficient material characterization process with scattering data.
\end{abstract}

\maketitle

\section{\label{sec:level1}Introduction}

Pair distribution function (PDF) analysis uses data from total X-ray or neutron synchrotron scattering experiments to quantitatively characterize short-range and long-range atomic structure.~\cite{egami2003underneath} Unlike conventional X-ray diffraction, which only measures the Bragg peaks that result from the atomic periodicity, PDF analysis also incorporates diffuse scattering, which results from disorder, allowing for structural characterization without assuming periodicity.~\cite{egami2003underneath,billinge2004beyond,nakamura2017unlocking} PDF analysis is thus applicable to crystalline, nanostructured, defected, and amorphous materials.~\cite{billinge2004beyond}

Advances in experimental techniques have increased the acquisition rates of PDF data. Advances in PDF analysis, however, are lagging. An experimental PDF is analyzed by adjusting the parameters of an assumed structure model, such as the lattice constant(s), atomic positions, and grain/particle size.  A refined PDF is obtained by minimizing the difference between the PDF of the structure model and the experimental PDF.~\cite{yang2020structure} This process is implemented in PDFgui,~\cite{farrow2007pdffit2} DiffPy-cmi,~\cite{juhas2015complex} and TOPAS.~\cite{coelho2015fast} Selecting the starting atomic structure(s) is a challenge and information about the sample (e.g.,  the crystal phase) is required to achieve satisfactory results. PDF analysis thus typically requires manual trial-and-error refinement of multiple  structure models.~\cite{banerjee2020cluster} 

One avenue to accelerating PDF analysis is to automate the refinement of a large number of structures that are pulled from a materials database or generated automatically.~\cite{yang2020structure,banerjee2020cluster} Such approaches are efficient when the composition and/or crystal structure are unknown. They may not be sufficient, however, to identify detailed structural information when the material and phase are known. For example, in defected materials and/or nanoparticles, there can be atomic displacements away from the perfect bulk structure due to defects and/or surfaces.~\cite{billinge2004beyond}

Data-driven methods, including machine learning, have emerged as a means to improve the efficiency of interpreting PDF data. Instead of independently performing PDF analysis on each member of a set of samples, data-driven methods attempt to learn the underlying patterns of the full PDF dataset. Principal component analysis (PCA) and non-negative matrix factorization (NMF), both unsupervised methods, are capable of extracting the PDF signals of the constituents of multi-phase and/or multi-component systems. PCA was applied by Chapman et al. to evaluate the evolution of the phase distribution during the nucleation and growth of zeolite-supported silver nanoparticles.~\cite{chapman2015applications} Li et al. used PCA to identify the chemical species involved in the discharge reaction of CoF$_2$.~\cite{li2018atomic} Liu et al. applied NMF to extract physically-interpretable components from PDF data of the lithiation of an RuO$_2$ electrode.~\cite{liu2021validation} Beyond PCA and NMF, machine learning models have also been used to study PDFs. Liu et al. built a convolutional neural network (CNN) to determine the space group of a material given its experimental X-ray PDF.~\cite{liu2019using} Anker et al. developed a conditional variational autoencoder to characterize the atomic structure of elemental metallic nanoparticles from PDF data.~\cite{anker2020characterising} 

\begin{figure*}
\centering
\includegraphics[scale=0.5]{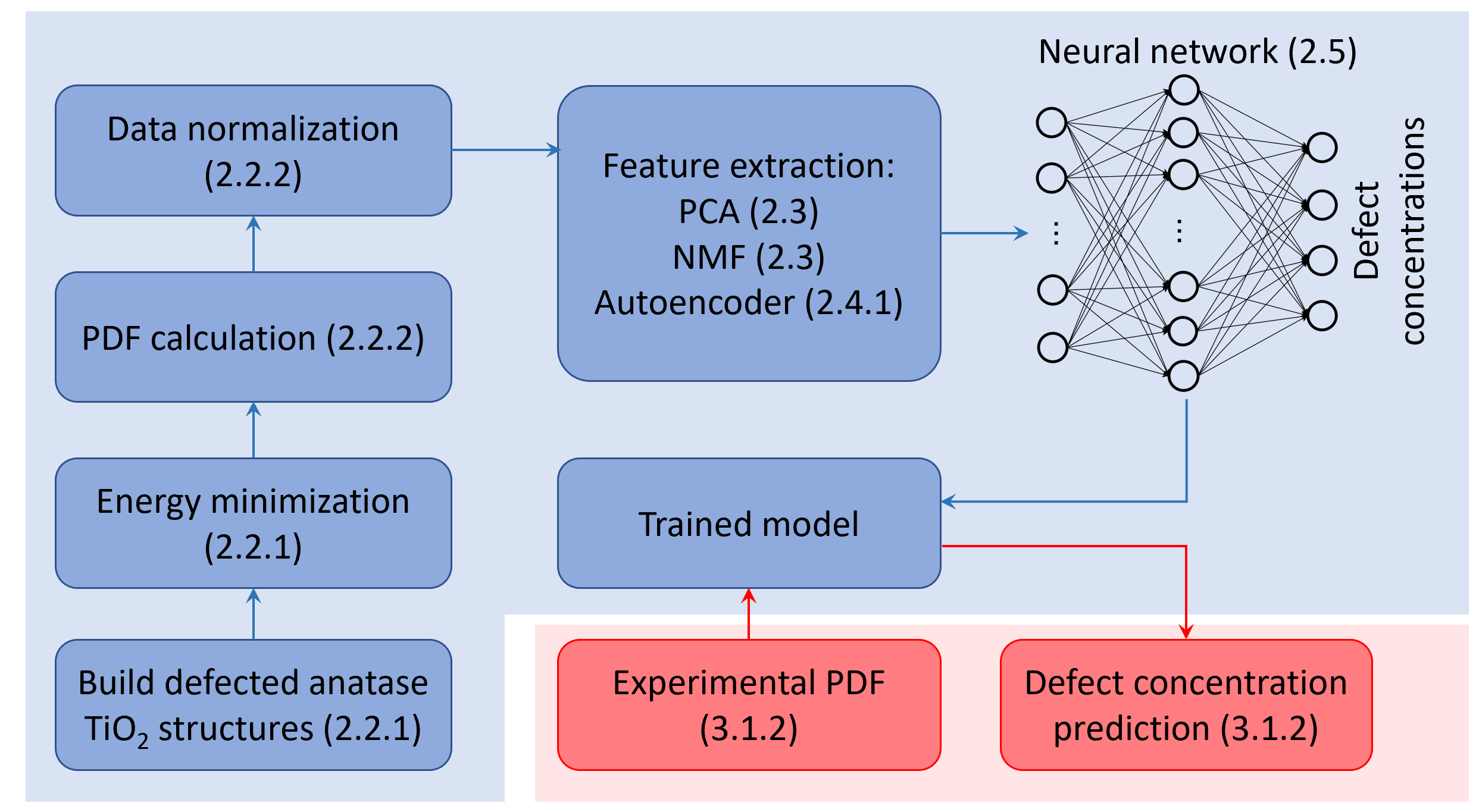}
\caption{Machine learning pipeline for predicting defect types and concentrations. The blue shaded area contains the data preprocessing and the training of the ML model with simulated PDF data. The red shaded area contains the application of the model to an experimental PDF.} 
\label{fig:flow}

\end{figure*}

Herein, we present a machine learning pipeline, shown in Fig.~\ref{fig:flow}, that combines: (i) unsupervised learning of the features embedded in PDF data, and (ii) supervised learning models that predict the types and concentrations of point defects. The pipeline is the applied to an anatase titanium dioxide (TiO$_2$) thin film synthesized with a sol-gel method.~\cite{nakamura2017unlocking} The blue shaded area in Fig.~\ref{fig:flow} contains the data preprocessing and the training of the machine learning model using simulated PDFs. The red shaded area contains the application of the model to an experimental PDF. 

The paper is organized as follows. An overview of the PDF refinement is provided in Sec.~\ref{pdf}. The preparation of the training/testing dataset of potential structures, which are built from atomistic simulations, is described in Sec~\ref{stru gen} and the PDF calculation is described in Sec.~\ref{pdf calc}. Due to the high dimensionality of each PDF (850 data points), using the unprocessed data to predict the defect concentration is inefficient algorithmically and computationally, and may introduce noise and cause overfitting.~\cite{liu2016overfitting} Feature extraction methods, which are described in Sec.~\ref{algorithms}, are hence required to reduce the dimensionality and discover patterns in the PDF data. Linear algorithms (PCA, NMF) and nonlinear algorithms (autoencoder, CNN) for feature extraction are compared by using their output to train a neural network to predict the defect types and concentrations in experimentally measured PDFs. 
The performance of the models is presented in Sec.~\ref{performance}, where we find that linear feature extraction models tend to provide better interpretability, while nonlinear ones provide better predictive ability. The learned features are discussed in Sec.~\ref{feature}.

\section{\label{methods}Methods}

\subsection{\label{pdf}Pair distribution function (PDF) analysis}

We use experimental PDFs of bulk anatase TiO$_2$ and an anatase TiO$_2$ thin film from a previous study~\cite{nakamura2017unlocking} to test the predictive ability of our models. The data acquisition was performed at the X-ray Powder Diffraction Beamline, 28-ID-2, at the National Synchrotron Light Source II at Brookhaven National Laboratory. For metal oxides with heavier metal elements, oxygen is relatively more sensitive to neutrons than to X-rays,~\cite{ren2018synchrotron} leading to stronger signals for oxygen and its defects. We use X-ray PDFs for TiO$_2$, however, because the negative neutron scattering length of Ti results in a cancellation between positive and negative peaks that makes some features hard to detect.~\cite{zhang2022pair}

The PDF, $G(r)$, gives a scaled probability of finding two atoms separated by a distance $r$.~\cite{egami2003underneath} Given a known structure model and ideal instrumental conditions, the PDF can be calculated by summing over all pairwise interatomic distances in a system with $N$ atoms from
\begin{equation}\label{eqn:pdf1}
G(r)=\frac{1}{Nr}\sum_i\sum_{j\neq i}\left [  \frac{b_ib_j}{\langle b \rangle^2}\delta \left ( r-r_{ij} \right )\right ]-4\pi r\rho_0.
\end{equation}
Here, $b_i$ is the scattering factor of atom $i$, $ \langle b \rangle$ is the compositional average scattering factor, $\delta()$ is the Dirac delta function, $r_{ij}$ is the distance between atoms $i$ and $j$, and $\rho_0$ is the atomic number density.

Two corrections are included in Eq.~\eqref{eqn:pdf1} to account for non-idealities. First, the peaks are broadened by a factor $\sigma$ given by~\cite{farrow2007pdffit2}
\begin{equation}\label{eqn:pdf2}
    \sigma = \sigma_{ij} \sqrt{1-\frac{\delta_1}{r_{ij}}-\frac{\delta_2}{r^2_{ij}}+Q^2_{broad}r^2_{ij}}.
\end{equation}
Here, $\sigma_{ij}$ is the uncorrelated peak width calculated from the atomic displacement parameters (ADPs), denoted by $U_{iso}$. The terms $\delta_1/r_{ij}$ and $\delta_2/r_{ij}$ correct for the effects of correlated atomic motion at short distances. Here, only $\delta_2/r_{ij}$ is considered to describe low-temperature behavior.~\cite{farrow2007pdffit2} $Q_{broad}$ is an instrument-dependent term.~\cite{farrow2007pdffit2}

Second, the presence of finite crystallite size tends to dampen the experimental PDF due to the size and shape of the coherently scattering domains.~\cite{kodama2006finite,howell2006pair} This effect is included by multiplying Eq.~\eqref{eqn:pdf1} by an envelope function that assumes spherical 
domains with diameter $d$ such that:
\begin{equation}\label{eqn:pdf3}
    G(r,d)=G(r)\left [ 1-\frac{3}{2}\frac{r}{d}+\frac{1}{2}\left ( \frac{r}{d} \right )^3 \right ] \theta (r,d),
\end{equation}
where $\theta$ is a step function with a value of unity for $r<d$ and zero otherwise.


\subsection{\label{data}Dataset preparation}
\subsubsection{\label{stru gen}Structure generation}

We built anatase TiO$_2$ structures with different types and concentrations of randomly-placed point defects. The simulation box of the perfect structure before introducing defects contains $6\times6\times9$ unit cells with 1944 atoms. Periodic boundary conditions are applied in all three directions. Combinations of common point defects (i.e., vacancies and interstitials) are considered for both the Ti and O atoms. Each structure has a label of four integers: $V_{Ti}$, $V_O$, $Ti_i$, and $O_i$, which are the numbers of Ti vacancies, O vacancies, Ti interstitials, and O interstitials. The number of each defect type takes a value $0$ \textendash~$80$, corresponding to concentrations of 0 \textendash~4.1\% (based on the number of atoms in the perfect structure). We built 800 structures. Among them, 480 contain all defect types, 160 have one defect type with zero concentration, 120 have two defect types with zero concentration, and 40 have three defect types with zero concentration. 

The box size and atomic positions in each structure was relaxed using energy minimization allowing with a conjugate gradient algorithm in the Large-scale Atomic/Molecular Massively Parallel Simulator (LAMMPS)~\cite{thompson2022lammps} until the relative energy change is smaller than $10^{-10}$ and the force in any direction is smaller than $10^{-10}$ eV/\AA. The second moment tight-binding charge equilibrium (SMTB-Q) variable charge potential~\cite{tetot2008tight,maras2015improved} is used due to its robustness in modeling defected TiO$_2$.~\cite{zhang2022pair} The electrostatic interaction was calculated using the Wolf summation method.~\cite{tetot2008tight, wolf1999exact} The charge on each atom was updated every relaxation step using the charge equilibrium scheme developed by Rappé and Goddard.~\cite{rappe1991charge} More details about the simulation setup and benchmarking of the potential can be found in our prior work.~\cite{zhang2022pair}

\subsubsection{\label{pdf calc}PDF calculation}

The PDFs of the relaxed atomic structures were calculated between 1.5~\AA~to 10 \AA~ in 0.01 \AA~ increments from Eqs.~\eqref{eqn:pdf1} \textendash~\eqref{eqn:pdf3} using DiffPy-cmi,~\cite{juhas2015complex} resulting in 850 data points for each PDF. $Q_{broad}$ is taken to be 0.0143 \AA$^{-1}$ from the experimental measurement on the TiO$_2$ samples to be analyzed.~\cite{nakamura2017unlocking} The calculated PDFs depend on the refineable parameters, which include a scaling factor, the lattice parameters, the ADPs, the low-$r$ peak sharpening coefficient $\delta_2$, and the spherical particle size $d$. Of these parameters, we find that the scale factor and spherical particle size most affect the peak heights. Our objective is to predict the defect concentration of experimental PDFs, where the signal strength can vary due to experimental conditions. As such, we eliminated the effect of the scale factor by normalizing each PDF by~\cite{liu2019using}
\begin{equation}
{\widetilde{G}(r)} = \frac{G(r) - \textrm{min}(G)}{\textrm{max}(G) - \textrm{min}(G) },
\end{equation}
such that $0 \leq \widetilde{G}(r) \leq 1$. $\delta_2$ and $d$ are then randomly selected from a set of physically-reasonable values. This process is repeated four times for each of the 800 structures. We thus obtain 3200 PDFs, with each structure having four PDFs calculated from four sets of parameters. The parameters used for the PDF calculations are listed in Table~\ref{tab:240 structures}.

\medskip
\begin{table}
\caption{PDF calculation parameters}
\label{tab:240 structures}
\begin{center}
\begin{tabular}{lc}
\toprule
$Q_{broad}$ (\AA$^{-1}$)    & 0.0143~\cite{nakamura2017unlocking}\\
$U_{iso,Ti}$ (\AA$^{2}$)        & 0.0046 ~\cite{zhang2022pair}\\
$U_{iso,O}$ (\AA$^{2}$)    &  0.0167 ~\cite{zhang2022pair}\\
$\delta_2$ (\AA$^{2}$)     & 1.0, 1.5, 2.0, 2.5, 3.0\\
$d$ (\AA)            & 20, 40, 60, 10$^8$ \\ \hline \hline
\end{tabular}
\end{center}

\end{table}

\subsection{\label{algorithms}Linear feature extraction: Principal component analysis (PCA) and non-negative matrix factorization (NMF)}

PCA and NMF are techniques where a matrix $\mathbf{V}$ is factorized into two matrices $\mathbf{W}$ and $\mathbf{H}$, such that its elements are
\begin{equation}\label{Eq_nmf}
 V_{i\mu}\approx (WH)_{i\mu}=\sum_{a=1}^{p}W_{ia}H_{a\mu}.
\end{equation}
For our PDF dataset, $\mathbf{V}$ is an $n \times m$ matrix of size $850 \times 3200$, where each column contains a PDF. $\mathbf{W}$ and $\mathbf{H}$ have dimensions of $n \times p$ and $p \times m$. The parameter $p$ is chosen based on the desired number of basis PDFs, which are contained in the columns of $\mathbf{W}$. Each column of $\mathbf{H}$ contains the weights in a one-to-one correspondence with a basis PDF in $\mathbf{W}$.~\cite{lee1999learning} The features to be used as the inputs to the neural network are the columns in the $\mathbf{H}$ matrix, with each one having dimension $p \times 1$. We chose $p = 9$ for PCA and $p=16$ for NMF so that 99.5\% of the variance is explained. 

$\mathbf{W}$ and $\mathbf{H}$ are solved as a optimization problem using the objective function $\left \| \mathbf{V-WH} \right \|_F$, where $\left \| \cdot \right \|_F$ indicates the Frobenius norm. In PCA, the columns of $\mathbf{W}$ are constrained to be orthonormal and the rows of $\mathbf{H}$ are constrained to be orthogonal,~\cite{liu2021validation} such that a unique solution is guaranteed.~\cite{liao2003network} In NMF, the elements of $\mathbf{V}$, $\mathbf{W}$, and $\mathbf{H}$ are constrained to be non-negative. Because this problem is non-convex, a unique solution is not guaranteed.~\cite{huang2013non} As such, we employ an initialization scheme called non-negative double singular value decomposition, which rapidly reduces the approximation error to a value that is potentially lower than that from a random initialization.~\cite{boutsidis2008svd} More importantly, we observe that it results in smoother PDF components compared to other initialization schemes. We perform PCA and NMF using the Scikit-Learn package.~\cite{pedregosa2011scikit}

\subsection{Non-linear feature extraction: Autoencoder}
 
An autoencoder is an unsupervised learning technique that leverages neural networks for the task of representation learning.~\cite{Goodfellow-et-al-2016} A neural network architecture is constructed with a bottleneck that forces a compressed knowledge representation of the original input. Here, the inputs are the simulated PDFs. The extracted features are embedded in the bottleneck layer. The single-layer autoencoder we constructed is shown in Fig. S1 in the Supplemental Material.~\cite{SM} A softplus activation function is applied at the hidden layer to ensure the positivity of the latent space $\mathbf{H}$. This autoencoder is similar to the NMF algorithm in that $\mathbf{H}$ contains the weights in a one-to-one correspondence with a basis PDF in the decoder weight matrix $\mathbf{W}$.~\cite{smaragdis2017neural} The difference is that there is no non-negativity constraint applied on the weights $\mathbf{W}$ in the autoencoder. The product of $\mathbf{W}$ and $\mathbf{H}$ approximates the original PDFs. We consider two initialization schemes. One is to randomly initialize the decoder weights (denoted as AE-rand) and the other is to initialize five of the weights to be the PDF signals of the perfect phase and the four types of defects, which are defined in Sec.~\ref{feature} (AE-phys). The autoencoders are implemented using the Pytorch package.~\cite{paszke2017automatic}

We also built a one-dimensional convolutional neural network that directly maps the PDFs to the defect concentrations, which is described in Sec. S2 of the Supplemental Material.

\subsection{\label{NN}Supervised learning of defect concentrations: Neural networks}

We trained neural networks using the $\mathbf{H}$ matrices obtained from PCA, NMF, AE-rand, and AE-phys as inputs and the numbers of defect as labels. We tuned the hyperparameters of the neural network with the random search method using five-fold cross validation on the simulated TiO$_2$ PDFs. The NN architecture with the highest averaged test score from the cross validation has two 50-neuron hidden layers that use the hyperbolic tangent activation function. The output layer has four neurons for the four defect types: $V_{Ti}$, $Ti_i$, $V_O$, and $O_i$. A scaled sigmoid activation function is used to bound the output values into the range of 0 to 80. The neural network is implemented using the Pytorch package.~\cite{paszke2017automatic}

\section{Results and discussions}

\subsection{\label{performance}Model performance}

\subsubsection{Training and testing}

\begin{figure*}
    \centering
    \includegraphics[scale=0.45]{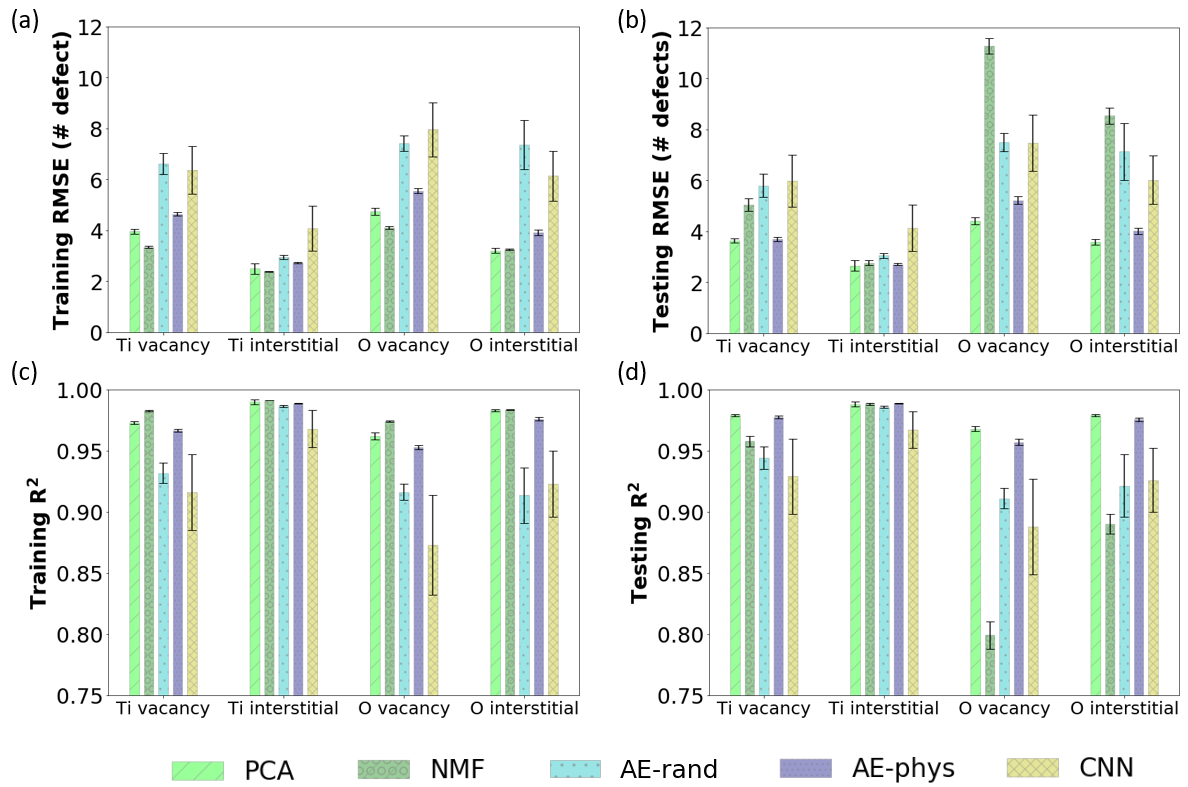}
    \caption{RMSE of (a) training data and (b) testing data, and the $R^2$ scores of (c) training data and (d) testing data, for the four defect types ($V_{Ti}$, $Ti_i$, $V_O$, and $O_i$) using five models. The bar graphs show the mean and the standard deviation for each type of model trained 50 times with different random seeds.}
    \label{fig:rmse} 
\end{figure*}

\begin{figure}
    \centering
    \includegraphics[scale=0.6]{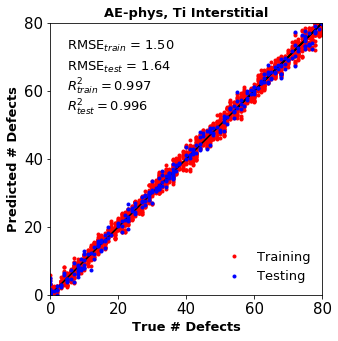}
    \caption{Ground truth of number of defects versus the prediction by AE-phys for the
    Ti interstitial. The red dots are training data and the blue dots are testing data.}
    \label{fig:parity} 
\end{figure}

As presented in Sec.~\ref{methods}, five models: PCA, NMF, AE-rand, AE-phys, and CNN are used for feature extraction. Each of the five models was trained on 80\% of the 3200 simulated PDFs of defected structures and tested on the remaining 20\%. This process was repeated 50 times for each model using a randomized train/test split and initialization of the neural networks. The autoencoder and CNN models have additional randomization due to the initial weights. The average root-mean-square error (RMSE) and its standard deviation for the predicted number of defects compared against the ground truth number of defects for the training and testing processes are shown in Figs.~\ref{fig:rmse}(a) and \ref{fig:rmse}(b). The  $R^2$ scores and their standard deviations are shown in Figs.~\ref{fig:rmse}(c) and \ref{fig:rmse}(d). All feature extraction methods result in a training RMSE lower than 8 defects and testing RMSE lower than 12 defects for all four defect types. Recall that the structures contain 0 to 80 defects of each type. The Ti interstitial RMSEs are the lowest for the training and testing sets for all models. As will be discussed in Sec.~\ref{feature}, the feature associated with Ti interstitials is the easiest to be separated. The $R^2$ scores for the training and testing sets are all above 0.85, except for the O vacancy predicted using NMF on the test data, which also has the largest testing RMSE. The parity plot for the Ti interstitial using AE-phys is shown in Fig.~\ref{fig:parity} as a demonstration of the defect concentration prediction.

With the same ratio of the variance explained (99.5\%), using NMF (dark green) to extract the features results in a smaller training RMSE compared to PCA (light green) for all defect types. The testing error for NMF is larger for O vacancies and O interstitials compared to their training error, suggesting a potential overfitting. Hyperparameter tuning for the neural network and lowering the number of components in the NMF, however, do not reduce the overfitting. Using AE-phys (dark blue) reduces both the training and testing RMSEs compared to the AE-rand (light blue). The standard deviations of RMSE and $R^2$ for CNN are consistently larger than for the other models. This result could be due to the much larger number of parameters in the CNN compared to the other models.~\cite{rawat2017deep}

\subsubsection{\label{exp_pdf}Application to experimental PDFs}

We applied the trained models to experimental PDFs of defected and perfect anatase TiO$_2$.~\cite{nakamura2017unlocking} The ground truths are [$V_{Ti}$, $Ti_i$, $V_O$, $O_i$] = [54($\pm$5), 34($\pm$5), 1($\pm$2), 9($\pm$6)] for the defected structure obtained via an iterative search in our previous work,~\cite{zhang2022pair} and [$V_{Ti}$, $Ti_i$, $V_O$, $O_i$] = [0, 0, 0, 0] for the perfect structure. 

\begin{figure}
    \centering
    \includegraphics[scale=0.45]{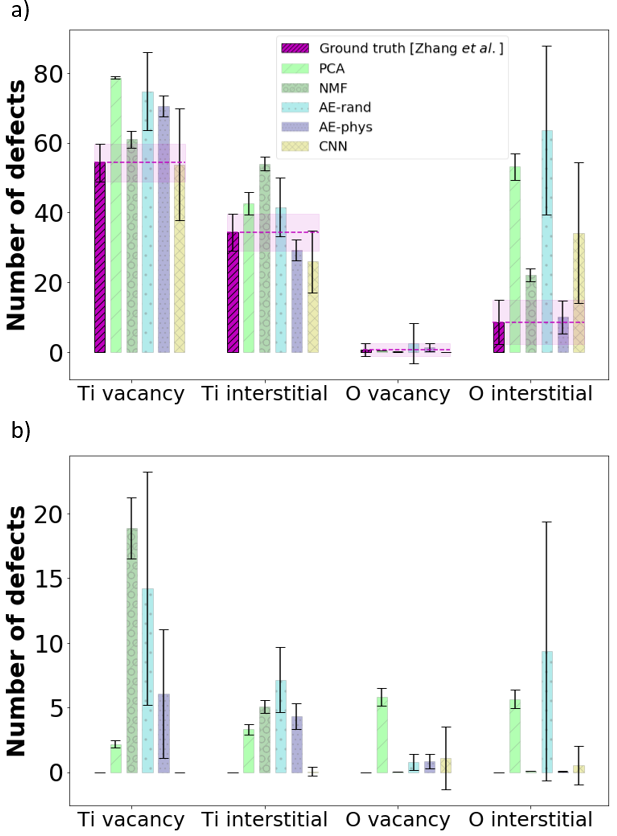}
    \caption{Predicted number of defects in an initially perfect 1944-atom cell for experimental TiO$_2$ PDFs using PCA, NMF, AE-rand, AE-phys, and 1D-CNN along with the ground truth for (a) a defected anatase TiO$_2$ sample from our previous work and (b) a perfect anatase TiO$_2$ sample. The bar graphs show the mean and the standard deviation for each type of model trained 50 times with different random seeds.}
    \label{fig:pred} 

\end{figure}

The predictions of the five models and the ground truth values are shown in Figs.~\ref{fig:pred}(a) and \ref{fig:pred}(b). The bar heights and error bars of the predictions are the means and standard deviations obtained from the 50 trained models. For defected TiO$_2$ [Fig.~\ref{fig:pred}(a)], all models predict a minimal number of O vacancies (smaller than 3), which is consistent with the ground truth value of ($1 \pm 2$). All models tend to overpredict the number of defects for the other three types, notably for O interstitials.
NMF performs better than PCA except for Ti interstitials. AE-rand overpredicts the number of Ti defects and the predicted number of O interstitials is seven times higher than the ground truth with a large  standard deviation. The performance of the autoencoder is significantly improved by the physics based initialization in AE-phys. The CNN gives good predictions on Ti vacancies, Ti interstitials, and O vacancies, but overpredicts on O interstitials and has a large standard deviation across all models.

The predictions for the perfect TiO$_2$ are shown in Fig.~\ref{fig:pred}(b). NMF predicts a high value for the Ti vacancy concentration, but lower O defects compared to PCA. AE-rand has the largest standard deviation, while the physics-based initialization of AE-phys again improves the prediction. 

Overall, AE-phys leads to the most accurate predictions on the experimental PDFs with relatively small variance. As noted above, the physical knowledge used in AE-phys contributes largely to improving the performance compared to AE-rand. Furthermore, nonlinear feature extraction methods perform better than linear methods. One explanation for this finding is that the PDF signals of the defects may be embedded in a more complicated way than the PDF signals of different materials in a mixture or solution, where PCA and NMF have previously been successful.~\cite{chapman2015applications, liu2021validation, li2018atomic}

To assess the AE-phys predictions from Fig.~\ref{fig:pred}(a), we created five random structures with the mean number of defects of each type. These structures were relaxed using the SMTB-Q potential by energy minimization. PDF refinement was then performed for each structure against the experimental PDF. The quality of the fit was quantified by $R_{\mathrm{w}}=\sqrt{\frac{\sum_{n}\left(G_{\mathrm{obs}, n}-G_{\mathrm{calc}, n}\right)^{2}}{\sum_{n} G_{\mathrm{obs}, n}^{2}}}$,~\cite{egami2003underneath}
where $G_{\mathrm{obs}, n}$ is the $n$th point on the experimentally measured PDF and $G_{\mathrm{calc}, n}$ is the $n$th point on the refined PDF. The average $R_w$ is $0.268 \pm 0.002$, which is in good agreement with that calculated using our previous brute force search approach of $0.266 \pm 0.002$.~\cite{zhang2022pair}

\subsection{\label{feature}Feature learning}

We now explore the possibility that the features extracted from PCA, NMF, AE-rand, and AE-phys can be interpreted in terms of the perfect structure and the four point defect types. To do so, the PDF signal of each point defect type $x$ is defined as $G^x = G^x_{30} - G_{perfect}$, where $G^x_{30}$ is the PDF of a structure with 30 defects of type $x$ and $G_{perfect}$ is the PDF of the perfect structure. To determine if an extracted feature corresponds to one of the five PDF physical components ($G_{perfect}$ and the four $G^x$), we calculated the Pearson correlation coefficient (PCC, $r_p$) between all pairs of extracted features and physical components for each of the four methods. 

The results are provided in Table~\ref{tab:rp}, where $i$ is the index of the extracted feature having the largest PCC with a specific physical component. PCC values greater than 0.8, which correspond to a strong correlation,~\cite{akoglu2018user} are bolded. We observe that the prediction accuracy for a specific defect concentration (Fig.~\ref{fig:pred}) does not strongly depend on the degree of correlation between the extracted features and that physical component. The extracted features and physical components for the different feature extraction methods are shown Figs. S3-S4 of the Supplemental Material.~\cite{SM}

\medskip
\begin{table}
\caption{Largest Pearson correlation coefficient ($r_p$) between the extracted features by different models and the physical components. $i$ is the index of the extracted feature having the largest $r_p$ with a certain physical components.}
\label{tab:rp} 
\begin{tabular}{lcc|cc|cc|cc|cc}
\toprule
 & \multicolumn{2}{c}{\bf Perfect}&\multicolumn{2}{c}{$\bf V_{Ti}$}&\multicolumn{2}{c}{ $\bf Ti_{i}$}&\multicolumn{2}{c}{$\bf V_{O}$}&\multicolumn{2}{c}{$\bf O_{i}$}\\\cline{2-11}
 &$i$ &$r_p$ &$i$ &$r_p$ &$i$ &$r_p$ &$i$ &$r_p$ &$i$ &$r_p$\\\hline
PCA &1 &0.60 &1 &-0.68 &2 &\bf -0.86 &4 &0.75 &7 &-0.54  \\
NMF &1 &\bf 0.96 &2 &0.68 &3 &\bf 0.94 &4 &0.66 &9 &0.58  \\
AE-rand &6 &0.53 &6 &-0.51 &1 &-0.70 &10 &0.52 &10 &0.37 \\
AE-phys &1 &\bf 0.99 &5 &0.63 &3 &\bf 0.95 &5 &\bf 0.87 &5 &0.47  \\ \hline \hline
\end{tabular}
\end{table}


AE-phys extracts three features with strong correlation to the physical components, while NMF, PCA, and AE extract two, one, and zero strongly-correlated features, respectively. The strong performance of AE-phys compared to AE-rand is a result of its initialization. Compared with PCA and NMF, AE-phys may perform better because it only applies the non-negativity constraint on the latent space by the Softplux activation function, but nothing on the basis PDFs (i.e., the decoder weights). NMF may perform better than PCA because it limits the component weights to be positive and does not apply an orthogonality constraint.

The link between the extracted features and the physical components is moderate. Consistent with the discussion in Sec.~\ref{exp_pdf}, this finding suggests that while the PDFs of the components or phases in a mixture or solution can be separated, the impact of atomic-scale features on the PDF is more difficult to elucidate.

\section{Conclusion}
We demonstrated a machine learning pipeline to predict the defect types and their concentrations in experimentally-measured PDFs of TiO$_2$. A dataset of defected structures was built and then relaxed using energy minimization. PCA, NMF, AE-rand, AE-phys, and CNN were used to extract features from the calculated PDFs that were then fed to a neutral network for defect concentration prediction. 

The RMSEs for the simulated PDFs are lower than 8 defects for the training set and lower than 12 defects for the testing set [Figs.~\ref{fig:rmse}(a) and \ref{fig:rmse}(b)]. When applying the trained models to the experimental PDFs, AE-phys generates the most accurate results with relatively small variance [Figs.~\ref{fig:pred}(a) and \ref{fig:pred}(b)]. We explored the physical significance of the extracted features and connections to the predictive capabilities of the different models. We found that AE-phys extracts the most number of features with strong correlation to the physical components (Table~\ref{tab:rp}). The strong performance of AE-phys is a result of its initialization and relaxed constraints. 
This machine learning pipeline can be expanded to other X-ray and neutron scattering data to facilitate a more efficient material characterization and discovery process.

\section*{Acknowledgements}

We thank Dr. Simon Billinge and Dr. Aarti Singh for helpful discussions. All authors acknowledge support from the Defense Advanced Research Projects Agency under award AIRA HR00111990030. B.R.-J. acknowledges support from the Army Research Office Young Investigator Program under contract number W911NF1710589.

\section*{Data availability}

The data that support the findings of this study are available from the corresponding author upon reasonable request.

\nocite{*}
\bibliography{aipsamp}

\end{document}



\title{Pair distribution function analysis for oxide defect identification through feature extraction and supervised learning}

\author{Shuyan Zhang$^1$}
\author{Jie Gong$^1$}%
\author{Daniel Xiao$^2$}
\author{B. Reeja Jayan$^1$}
\author{Alan J. H. McGaughey$^{1}$}\email{mcgaughey@cmu.edu.}
\affiliation{%
 $^1$Department of Mechanical Engineering, Carnegie Mellon University, \\Pittsburgh, Pennsylvania 15213, USA  
}%
\affiliation{%
 $^2$Department of Materials Science and Engineering, Carnegie Mellon University, \\Pittsburgh, Pennsylvania 15213, USA  
}%

\clearpage

\maketitle
\clearpage

\section{\label{sec:level1}Autoencoder structure}

\begin{figure}[h]
    \centering
    \includegraphics[scale=0.4]{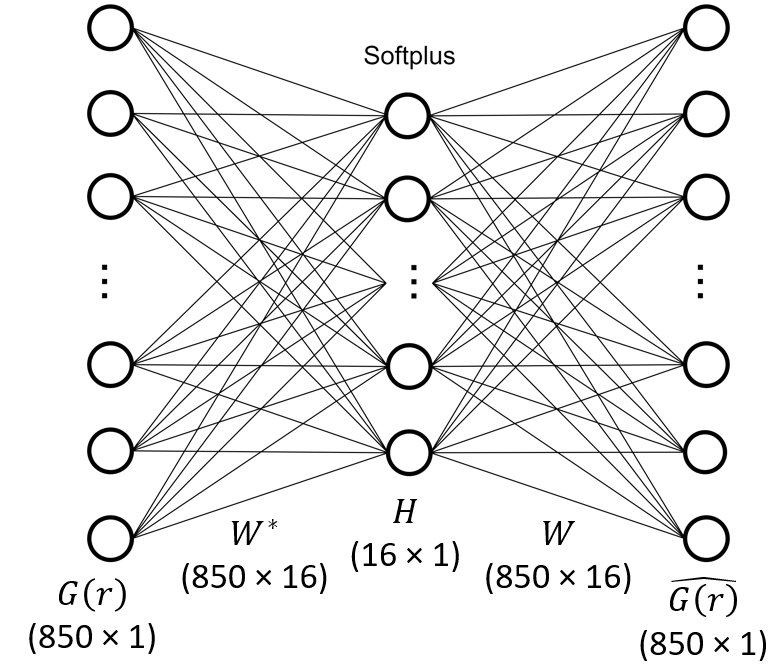}
    \caption{\label{fig:ae} Autoencoder used to extract features from the PDF dataset.}
\end{figure}

\section{\label{sec:level1}One-dimensional convolutional neural network}

A convolutional neural network (CNN) is a class of neural network that has a shared-weight architecture of convolution kernels that slide along input features and provide translation equivariant responses known as feature maps~\cite{kiranyaz20211d}. A CNN usually consists of an input layer, several hidden layers (e.g., convolution layers, pooling layers, and fully connected layers), and an output layer. One-dimensional CNNs (1D-CNNs) are CNNs with one-dimensional filters on the convolution layers, making them suitable for 1D signal extraction~\cite{kiranyaz20211d}. 

For the PDF dataset, the built-in feature extraction ability of a 1D-CNN makes it possible to extract information from adjacent peaks in a PDF. The convolution layers are effective for detecting local features within the kernel. The translational equivariant response from the convolution operation and the use of pooling layers helps to reduce the noise in PDFs calculated with different parameters. 

\begin{figure}[h]
    \centering
    \includegraphics[width=0.97\textwidth]{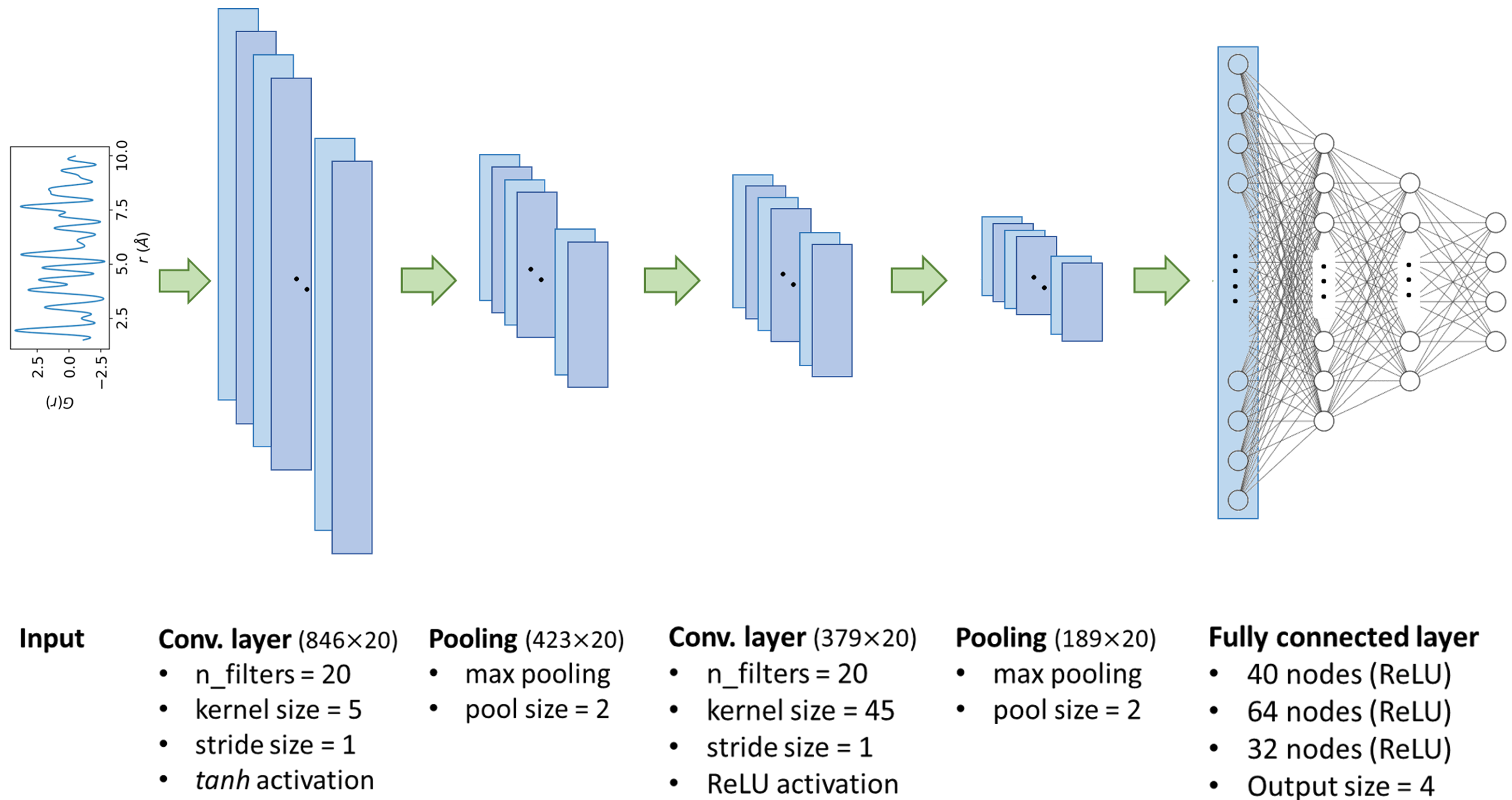}
    \caption{\label{fig:1DCNN} 1D-CNN architecture used in this study.}
\end{figure}

The architecture of our 1D-CNN is shown in Fig.~\ref{fig:1DCNN}. The hyperparameters were obtained through a search with the hyperband method~\cite{li2017hyperband}. The model is trained for 60 epochs using the Adam optimizer. The 1D-CNN method is self-contained compared with principal component analysis, non-negative matrix factorization, and an autoencoder (i.e., no additional neural network is needed). The 1D-CNN and the hyperband method are implemented with the Keras~\cite{chollet2015keras} library on the TensorFlow version 2.0 deep learning platform~\cite{tensorflow2015-whitepaper}. 

\clearpage

\section{\label{sec:level1}Extracted basis PDFs}

\begin{figure}[h]
    \centering
    \includegraphics[width=1\textwidth]{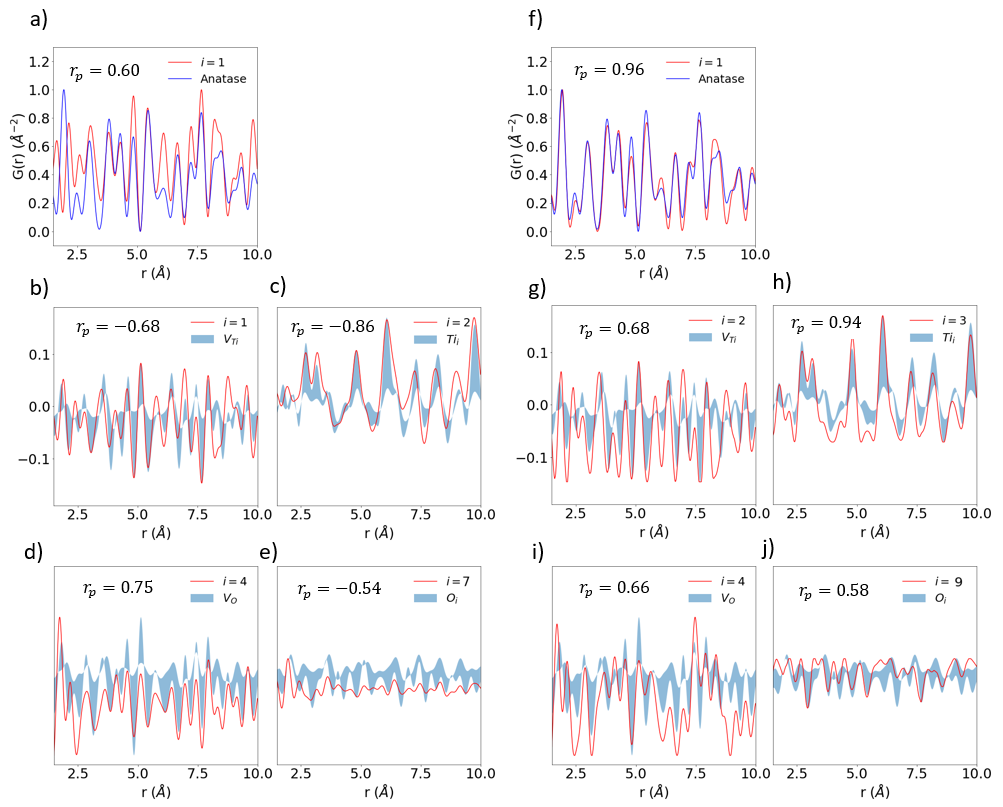}
    \caption{\label{fig:pca_nmf} Basis PDFs extracted (the red curves) from (a)-(e) PCA and (f)-(j) NMF compared with the physical components (blue curves).}
\end{figure}

The Pearson correlation coefficients (PCCs) reported in Table II of the main text correspond to the extracted features that provide the highest PCC with each of the five physical components.  These extracted feature and physical component pairs are plotted in Figs.~\ref{fig:pca_nmf}(a)-\ref{fig:pca_nmf}(e) for PCA, in \ref{fig:pca_nmf}(f)-\ref{fig:pca_nmf}(j) for NMF, in Figs.~\ref{fig:ae}(a)-\ref{fig:ae}(e) for the randomly-initialized autoencoder and in Figs. \ref{fig:ae}(f)-\ref{fig:ae}(j) for the physics-based initialized autoencoder shown. The physical components have a spread because they is calculated from the PDFs of structures that contain a range of concentrations for the same defect type. 

\clearpage

\begin{figure}[h]
    \centering
    \includegraphics[width=1\textwidth]{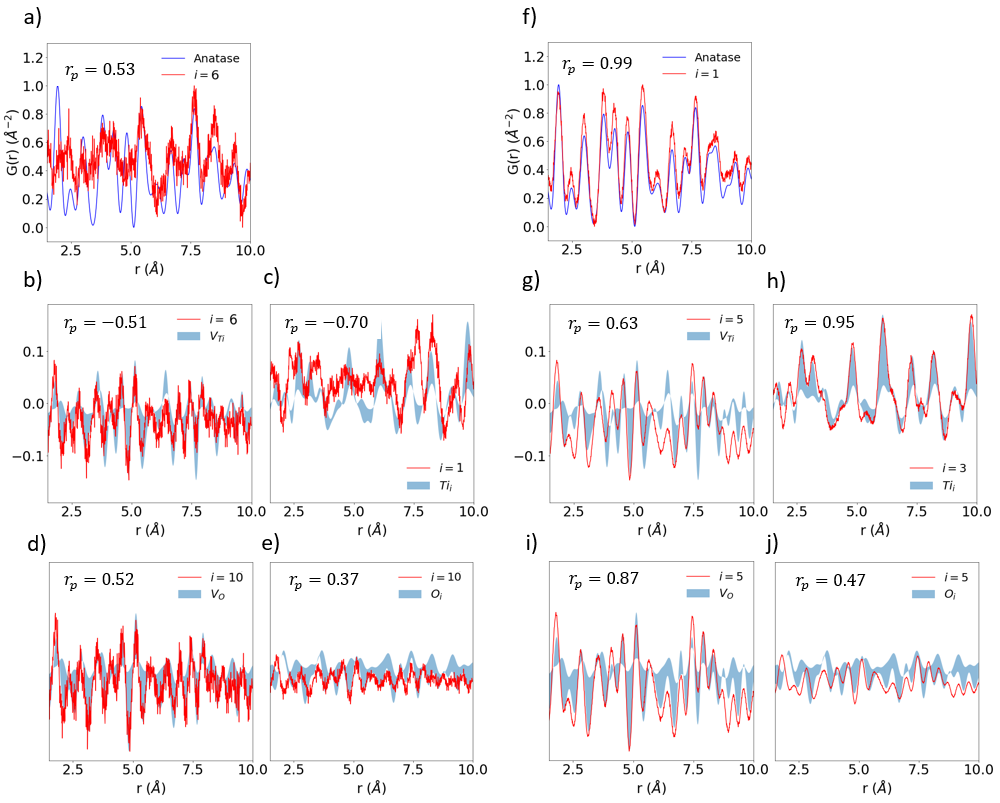}
    \caption{\label{fig:ae} Basis PDFs extracted (the red curves) from (a)-(e) randomly initialized autoencoder and (f)-(j) physics-based initialized autoencoder compared with the physical components (blue curves).}
\end{figure}
\clearpage

\nocite{*}
\bibliography{supplemental}